\newcommand{\ergs}{erg s$^{-1}$}
\documentclass{aa}  
\usepackage{graphicx}
\usepackage{txfonts}
\usepackage{comment}
\usepackage{multirow}
\usepackage{xcolor}
\def\kms{km~s$^{-1}$}

\begin{document} 
\title{A Newborn AGN in a Starforming Galaxy }
\titlerunning{Black hole ignition }
\authorrunning{P. Arévalo et al.}
   \author{P. Arévalo\inst{1,2}, E. López-Navas\inst{1,2}, M.L. Martínez-Aldama\inst{3,2}, P. Lira\inst{4,2}, S. Bernal\inst{1,2},  P. Sánchez-Sáez\inst{5}, M. Salvato\inst{7,8}, L. Hernández-García\inst{1,6}, C. Ricci\inst{9}, A. Merloni\inst{7}, M. Krumpe\inst{10}}
\institute{Instituto de F\'isica y Astronom\'ia, Universidad de Valpara\'iso, Gran Breta\~na 1111, Valpara\'iso, Chile\\
\email{patricia.arevalo@uv.cl}
\and Millennium Nucleus on Transversal Research and Technology to Explore Supermassive Black Holes (TITANS)
\and Astronomy Department, Universidad de Concepción, Casilla 160-C, Concepción 4030000, Chile
\and Departamento de Astronom\'ia, Universidad de Chile, Casilla 36D, Santiago, Chile
\and European Southern Observatory, Karl-Schwarzschild-Str. 2, 85748, Garching, Germany
\and Millennium Institute of Astrophysics (MAS),Monse\~nor S\'otero Sanz 100, Providencia, Santiago, Chile
\and Max-Planck-Institut für extraterrestrische Physik, Giessenbachstr. 1, 85748 Garching, Germany
\and Exzellenzcluster ORIGINS, Boltzmannstr. 2, 85748 Garching, Germany 
\and Instituto de Estudios Astrof\'isicos, Facultad de Ingenier\'ia y Ciencias, Universidad Diego Portales, Av. Ej\'ercito Libertador 441,
Santiago, Chile
\and Leibniz-Institut für Astrophysik Potsdam, An der Sternwarte 16, 14482
Potsdam, Germany
}

  \abstract
   {}
   {We report on the finding of a newborn AGN, i.e. current AGN activity in a galaxy previously classified as non-active, and characterize its evolution.}
   {Black hole ignition event candidates were selected from a parent sample of spectrally classified non-active galaxies (2.394.312 objects), that currently show optical flux variability indicative of a type I AGN, according to the ALeRCE light curve classifier. A second epoch spectrum for a sample of candidate newborn AGN were obtained with the SOAR telescope to search for new AGN features. }
   {We present spectral results for the most convincing case of new AGN activity, for a galaxy with a previous star-forming optical classification, where the second epoch spectrum shows the appearance of prominent, broad Balmer lines without significant changes in the narrow line flux ratios. Long term optical lightcurves show a steady increase in luminosity starting 1.5 years after the SDSS spectrum was taken and continuing for at least 7 years. MIR colors from the WISE catalog have also evolved from typical non active galaxy colors to AGN-like colors and recent X-ray flux detections confirm its AGN nature. }
   {}


   \maketitle
%

\section{Introduction}
Observations and models indicate that the fraction of active galaxies in the local Universe is about $\sim 10\%$ \citep{Schulze2010,Shankar2013,Sun2015}, which can be interpreted as a duty cycle, where 10\%\ of galaxies are active at any given time. Indirect evidence also suggests that activity varies by several orders of magnitude in time scales of $\Delta t \sim 10^4 - 10^7$ yrs, effectively turning the active nuclei on and off \citep[][and references therein]{Hickox2014, Ichikawa2019}.  Estimating this activation rate, or alternatively, how many times each galaxy has switched on and off, is important to constrain central black hole feeding mechanisms in galaxy evolution models. 

Newborn AGN, which involve a galaxy transitioning from a quiescent or star-forming (SF) state to a type I Active Galactic Nucleus (AGN), are exceptionally challenging to detect.  Part of the difficulty arises from the data available, since the largest spectroscopic survey, from SDSS, was originally shallow, it would target mostly galaxies that were bright at that time, and might have become dimmer, not the other way around. Among all the possible signatures of AGN activity, we will focus on optical spectral classification. This choice is justified by the availability of archival spectral data, and the ease of obtaining new spectra for a few sources. Moreover, AGN activity can be identified largely unambiguously through this approach. A key characteristic of Seyfert I galaxies and quasars, is the presence of broad emission lines in their optical spectra, with widths of thousands of \kms\ \citep{Baldwin1978}. Therefore, the identification of broad emission lines in a previously spectroscopically-classified quiescent galaxy could serve as compelling evidence for a black hole ignition event.

In the public archives, there are approximately two million galaxies with optical spectra showing no broad emission lines (or other features that identify them as AGN) obtained on average about a decade ago. Detecting ignition events within this vast dataset is possible if they occur more frequently than about 1/20,000,000 per year. Re-observing all of these galaxies, however, is impractical so alternative criteria for target selection are needed. Fortunately, there is another distinguishable characteristic of quasars and Seyfert I galaxies, which is their persistent and stochastic optical flux variability \citep[e.g.][]{ MacLeod10,Sanchez-Saez18}. Such variations are considerably rare in even type II AGN \citep{Lopez23}, and even rarer in quiescent galaxies as discussed below.

In this study, we present our first successful identification of a transition from a non-active to a type I AGN, accomplished by selecting targets based on optical variability and further validated by the appearance of broad, permitted emission lines in the optical spectrum.

\section{Selection Method}

We used the Automatic Learning for the Rapid Classification of Events (ALeRCE) light curve classifier (lcc) \footnote{Publicly available at \url{https://alerce.online}}\citep{Sanchez-Saez21, Forster21} to select galaxies which are currently varying as type I AGN. This classifier uses the alerts (i.e. a $5\sigma$ change in flux with respect to a previous epochs) of the Zwicky Transient Facility \citep[ZTF,][]{Masci19}  to characterize the variability of objects in the $g$ and $r$ bands. We cross matched this sample to a sample of probable quiescent galaxies with SDSS spectra and classifications in the DR16 \citep{Ahumada2020}, i.e. 2.300.000 objects with no detection of broad emission lines and no Seyfert or LINER classification according to the BPT \citep{baldwin1981classification} diagrams, or other AGN classification. This match produced our ignition candidate sample of only 86 objects, showing how rare  significant fluctuations are in non-active galaxies. From these, 18 galaxies were targeted for a confirmation optical spectrum, selected based on their visibility from Cerro Pachón, Chile, in the first semester, and on their brightness, with time awarded through CNTAC program SOAR2022A-010. The galaxy SDSS J080304.74+220734.0 (ZTF ID: ZTF20aaglfpy\footnote{\url{https://alerce.online/object/ZTF20aaglfpy}}, $z=0.1245$) showed the most significant change of activity and is presented in detail below.   

\section{Confirmation Spectrum}

A spectrum of ZTF20aaglfpy taken in MJD 52943 (2003-10-31, plate 1584) with the SDSS spectrograph led to a Galaxy classification. This spectrum showed no evidence for broad emission lines, and the ratios of its narrow emission lines corresponded to the star forming portions of the BPT diagnostics diagrams. We obtained a new optical spectrum for ZTF20aaglfpy with the Goodman High Throughput Spectrograph mounted on the SOAR telescope. Two exposures of 700 s each were taken with a 1$\arcsec$ slit, using grating SYZY 400, as well as lamp flats and arc exposures before and after the target. Observations were carried out on 2022-02-07.  The data were reduced using the SOAR pipeline\footnote{\url{https://github.com/soar-telescope/goodman_pipeline}} and the flux calibration was done using standard IRAF routines and stellar spectra obtained on the same night. Fig. \ref{fig:ZTF20aaglfpy_spectrum} shows the comparison of the old (SDSS) and new (SOAR) spectra for this source in the top panel.  Broad permitted lines have appeared in the new spectrum together with a bluer continuum, most evident in the  residuals section of the top panel.

\subsection{Spectral fitting and results}

We performed the fitting with pPXF \citep{2004PASP..116..138C,Cappellari2017}, including templates for the stellar contribution of the host galaxy, a composite powerlaw component for the AGN accretion disc, and two emission line templates of linked widths and velocity shifts for the lines, one for permitted and forbidden lines together and one for permitted lines only, although the width of H$\alpha$ was left free due to the evidence of possible smaller widths with respect to H$\beta$ \citep{reines2013}. The  stellar templates correspond to the population synthesis models of \citet{falcon2011updated} which are based on the spectra of the MILES library. For the AGN continuum, we used a linear combination of powerlaws with indices between -3 and +2 with steps on 0.1, with free normalizations.  FeII pseudo-continua are also included in the model. 

The archival SDSS spectrum was well fitted with a combination of stellar populations and one template of narrow emission lines. For the SOAR spectrum, we took the stellar template fitted to the SDSS spectrum, with a variable normalization and allowed all other parameters to vary. The resulting fit requires the addition of broad Balmer lines and a powerlaw of negative slope indicating the relevance of the accretion disk emission. The middle panels in Fig. \ref{fig:ZTF20aaglfpy_spectrum} show the spectral regions around H$\beta$--O[III] and H$\alpha$--NII--SII, for the old SDSS spectrum (left) and the new SOAR spectrum (right), plotted in black, together with the best-fitting model in cyan, after subtraction of the continuum. The broad  H$\beta$ and H$\alpha$ lines, marked in green, are only required in the new spectrum. 

The narrow emission lines fitted to each spectrum, plotted in orange in the same panel, have flux ratios as shown in the bottom row of panels. These plots correspond to the emission line diagnostics of \citet{baldwin1981classification}, with the limiting lines for star-forming, AGN and LINERs of \cite{kewley2001theoretical} and \citet{schawinsky2007}. The line dividing the SF and composite regions was taken from \cite{kewley2006host}. The symbols correspond to our fits to the archival SDSS spectrum (green circle), the new SOAR spectrum (red star) and an alternative fitting routine to the archival SDSS spectrum available from the RCSED$^2$ database \citep{Chilingarian2012,Chilingarian2017} {\footnote{\url{http://rcsed-dev.sai.msu.ru/}} 
(black square). In all cases the narrow line emission is consistent with star formation or composite between star formation and LINER activity.  Therefore this galaxy shows evidence of a broad line region typical of type I AGN, but a potential narrow line region close to the AGN is either absent of insufficiently illuminated by the nucleus to produce AGN-like line ratios, yet, pointing to no AGN activity for the last thousands of years. 

\begin{figure*}
    \centering
    \includegraphics[width=0.99\textwidth]{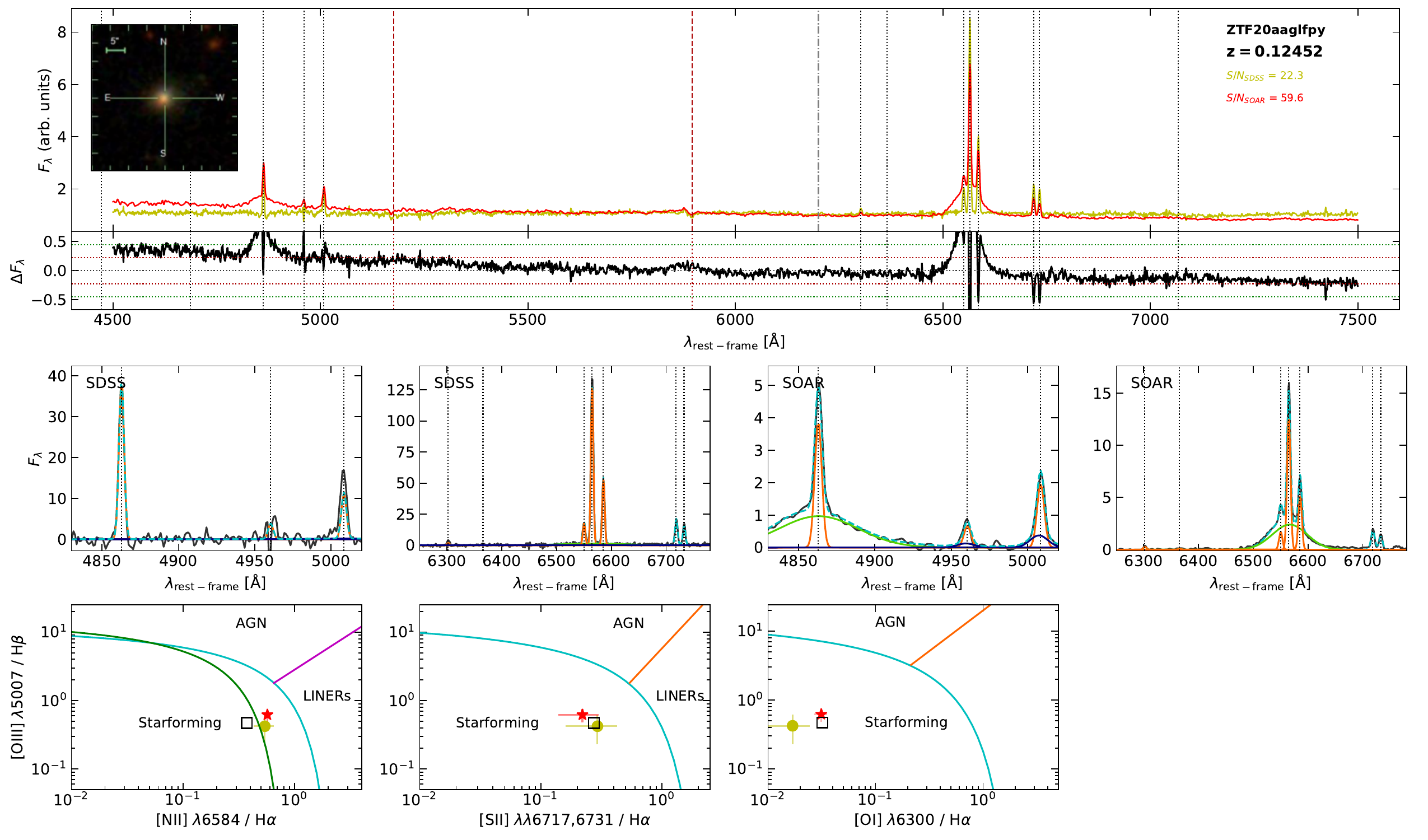}
    \caption{Top: original SDSS spectrum (yellow) and new SOAR spectrum (red) of ZTF20aaglfpy. Both spectra are normalized at 6200 \AA\ (vertical dot-dashed gray line). Vertical dotted lines indicate the main emission lines, while red dashed lines correspond to the Mg and Na absorption lines.  Inset panel shows the galaxy field of a $40\arcsec \times 40\arcsec$ region around the galaxy, taken from the SDSS database.  The residuals in this panel highlight the difference spectrum, with a blue slope and broad Balmer lines. Horizontal lines indicate the zero-level (black line), and the rms limits at $1\sigma$ (red line) and $2\sigma$ (green line), respectively. Middle panels: regions around H$\beta$ and H$\alpha$ for the old (left) and new (right) spectra, together with the fitted model, narrow emission lines in orange, broad emission lines in green, and blueshifted components for O[III] (blue).  Bottom panels: three BPT diagnostics diagrams, for different combinations of narrow emission line fluxes, the green circles represent our fits to the archival SDSS spectrum and red stars to the new SOAR spectrum. Black squares are the values derived for the archival SDSS spectrum presented in the RCSED$^2$ catalog.}
    \label{fig:ZTF20aaglfpy_spectrum}
\end{figure*}

\section{Long term flux evolution}
 The top panel in Figure \ref{fig:ZTF20aaglfpy_lcs} shows the long term optical flux evolution of ZTF20aaglfpy, where we combine data from the Catalina Real Time Survey \citep{Drake2009}(CTRS, black), with the ZTF forced photometry light curves in $g$ (green) and $r$ (red) bands, all averaged in 50-day bins and shifted vertically as stated in the figure legend. The CTRS data cover the period from 2005-04-09 to 2016-04-27 and the ZTF data from 2018-03-28 to 2023-04-25, therefore the SDSS spectrum, taken on 2003-10-31 predates the beginning of the light curves by 1.5 years, and the SOAR spectrum was taken during the ZTF monitoring, when the source was already showing variability typical of a type I AGN. 
CTRS data are acquired with an open CCD, without filters, so their magnitudes are not readily comparable to standard systems. Here we plot it shifted downward in the y-axis to acknowledge the fact that this lightcurve traces a broad wavelength range which includes the $g$ and $r$ bands used by ZTF. 
\begin{figure}
    \centering
    \includegraphics[width=0.52\textwidth]{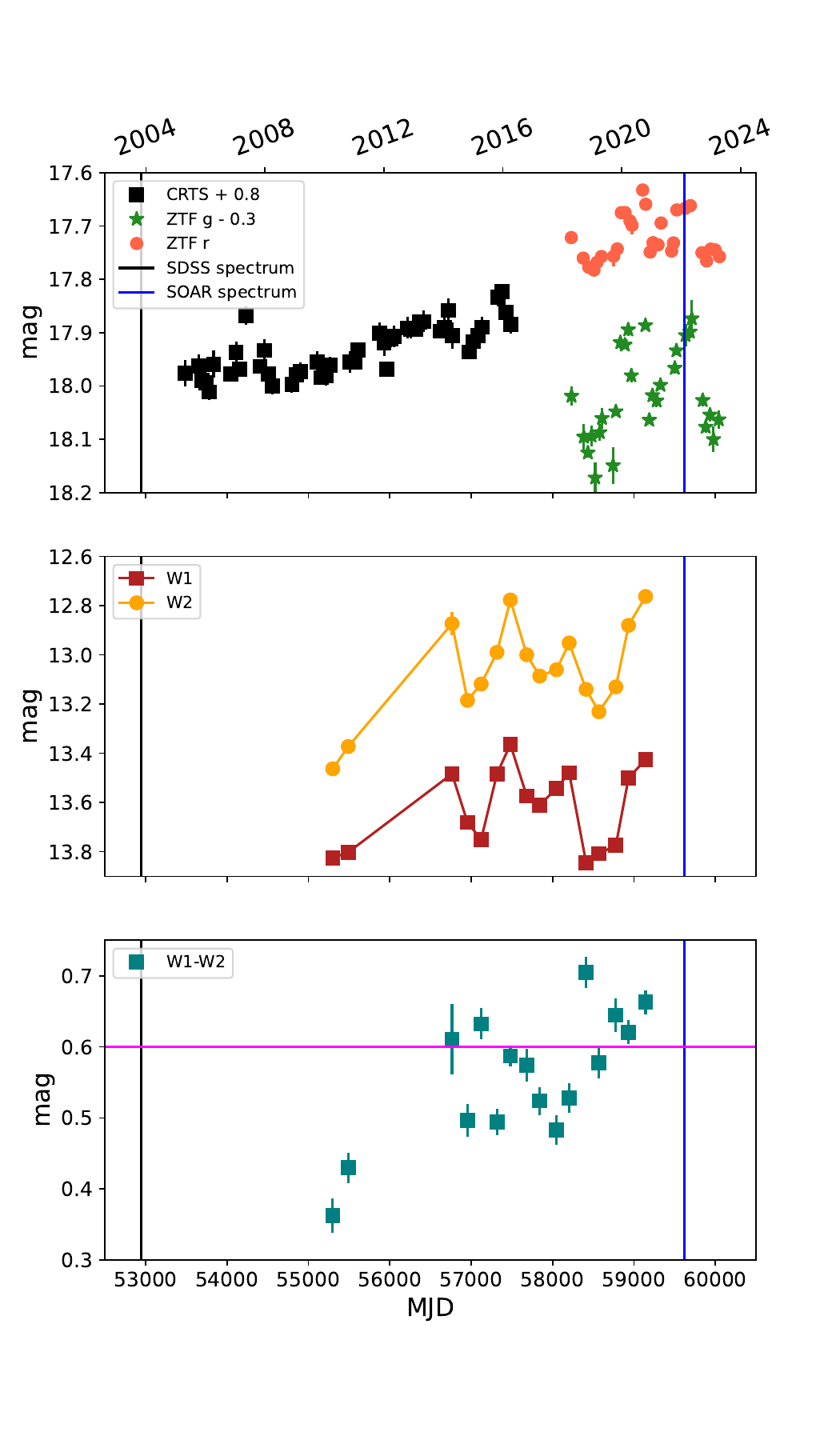}
    \caption{Top: Optical light curves of ZTF20aaglfpy as labeled in the legend. The CRTS has been shifted vertically by 0.8 mag and the ZTF g-band lightcurve by -0.3 mag to reduce the vertical scale. Middle: MIR flux in the WISE W1 and W2 filters from the unTimely catalog. Bottom: MIR colour evolution, based on the magnitudes in the middle panel. The horizontal line in the bottom panel marks the AGN-like region at W1-W2>0.6. All panels: The black vertical line marks the date of the SDSS spectrum and the blue line marks the date of the SOAR confirmation spectrum.  }
\label{fig:ZTF20aaglfpy_lcs}
\end{figure}

The CRTS light curve shows that the optical flux had been rising steadily from about MJD 55000, and all light curves show several more rapid flares by 0.1 (in CTRS and ZTF $r$ band) to 0.2 (in ZTF $g$ band) magnitudes starting from around MJD 57000. This long rise time and repeated flaring is inconsistent with other sources of optical variability such as SNe and TDEs, and point to the brightening of an AGN component. 
\subsection{Evolution of the Mid-Infrared flux}
 The middle panel in Figure \ref{fig:ZTF20aaglfpy_lcs} shows the MIR lightcurves of ZTF20aaglfpy in the \textit{Wide-field Infrared Survey Explorer} \citep[WISE][]{Wright2010} W1 (3.4 $\mu$m) and W2 (4.6 $\mu$m) filters, obtained by \citet{Meisner2023} from images co-added over periods of 6 months, i.e. the unTimely catalog. These lightcurves start around MJD 55000, close to the date when the CTRS optical lightcurve begins to rise steadily. The W1 and W2 lightcurves show significant, correlated variability over years timescales and their peaks and dips appear to line up with the optical fluctuations shown in the panel above \citep[see also][]{Brogan2023} . 

The MIR color W1-W2 can be used as a diagnostic of the origin of the MIR emission. For example, figure 9 in \citet{Comparat2020} shows the W1-W2 (Vega) color in the y-axis, for a variety of astronomical sources, where quiescent galaxies typically have values W1-W2<0.4 and type 1 AGN and quasars have mainly W1-W2 between 0.6 and 1.6. The authors propose a limit of W1-W2>0.7 to classify an object as AGN since very few stars and galaxies lie above this level, although many AGN show lower values.  The bottom panel of Fig. \ref{fig:ZTF20aaglfpy_lcs} shows the W1-W2 colour evolution of ZTF20aaglfpy, built from the unTimely lightcurves \citep{Meisner2023}. The colour evolves from typical of quiescent galaxies in the very beginning to colours more typical of type 1 AGN, i.e although only one point reaches the W1-W2>0.7 limit in the colour curve, all the final point go above W1-W2=0.6, placing this galaxy in AGN territory. 

\subsection{X-ray detections}
X-ray emission usually accompanies AGN activity but it does not represent an unambiguous criterion on its own, although large amplitude persistent X-ray variability does. Recent X-ray fluxes could be obtained for ZTF20aaglfpy from the eROSITA \citep[extended ROentgen Survey with an Imaging Telescope Array, ][]{2021erosita} All-Sky Surveys (eRASS1 to eRASS4). The data were processed with the eROSITA Standard Analysis Software System (eSASS, \citealt{2022brunner}) version c020, similar to that used for the eROSITA Data Release (Merloni et al. 2024, in press). 

The 0.2--2.3 keV X-ray fluxes from eROSITA, all obtained after the onset of optical variability, are summarized in Table \ref{tab:xrays}, showing variations of up to a factor 2. These variations point to an AGN rather than star formation origin, for at least part of the X-ray emission. Unfortunately there were no previous pointed observations with other X-ray telescopes, so we cannot confirm an overall brightening of the X-ray source over time. Upper limits are available from the XMM-Newton Slew Survey and from the ROSAT all sky survey, through the ESA multimission upper limit server.\footnote{http://xmmuls.esac.esa.int/upperlimitserver/}, which fall above the eROSITA detections. These upper limits, one taken even before the archival SDSS spectrum (ROSAT) and one during the optical/MIR brightening phase (XMM-Newton) at least confirm that the source was not much \emph{brighter} in X-rays then than it is now. 

\begin{table}[]
    \caption{X-ray flux measurements and archival $2\sigma$ upper limits of ZTF20aaglfpy.}
    \centering
    \begin{tabular}{|l|l|l|}
    \hline
    date &0.2-2.3 keV flux &Mission\\
MJD&[$10^{-13}$erg\,cm$^{-2}$\,s$^{-1}$]&\\
    \hline
    48163&<4.2& ROSAT all sky survey\\
    57116&<8.2& XMM-Newton slew survey\\
\hline
58961&$3.2\pm{0.7}$&\multirow{4}{1pt}{eROSITA}\\
59147&$1.9\pm{0.5}$&\\
59329&$2.6\pm{0.5}$&\\
59514&$1.7\pm{0.5}$&\\
      \hline

    \end{tabular}
    \label{tab:xrays}
\end{table}

\section{Characterization of the AGN in SDSS J080304.74+220734.0}
The current X-ray flux can be used to estimate the bolometric luminosity of the AGN. Assuming a luminosity distance of D=586.5 Mpc based on the redshift obtained from the optical spectrum z=0.1245, and a 2--10 keV flux of 2.15$\times 10^{-13}$ erg\,cm$^{-2}$\,s$^{-1}$ (calculated assuming a powerlaw spectrum with photon index $\Gamma=2$, and an average 0.2--2.3 kev flux of 2.5$\times 10^{-13}$ erg\,cm$^{-2}$\,s$^{-1}$), results in a 2--10 keV luminosity $L_{2-10}=9\times10^{42}$ erg\,s$^{-1}$. Using the hard X-ray bolometric correction for type I AGN of this luminosity of \citet{Duras2020}, $K_X=13.5$, results in a bolometric luminosity $L_{\rm bol,2-10}=1.2\times10^{44}$ erg\,s$^{-1}$. 

The bolometric luminosity can also be estimated from the optical continuum attributed to the AGN component. From the spectral analysis of the SOAR spectrum we estimated a luminosity at 5100 \AA\ of the powerlaw component of  log~$L_{5100}$=43.39$\pm$0.08 erg\,s$^{-1}$. Using the scaling relation of \citet{Shen08}, we obtain $L_{\rm{bol,5100}}= 2.3\times10^{44}$ erg\,s$^{-1}$. Considering that both scaling relations carry an uncertainty of about 0.3 dex and that the X-rays vary by a factor of about 2, the estimates of  $L_{\rm{bol}}$ from X-ray and optical bands are consistent. Therefore the relative flux between X-ray and optical bands is as expected for AGN of this luminosity.

The detection of the powerlaw component in the SOAR spectrum allows us to estimate the size of the broad line region through the Radius-Luminosity relation found by \citet{bentz13}, resulting in a size of 16.8 light days. The broad component of H$\beta$ has a width of $3852\pm73$ \kms. Following the description given by \citet{MejiaRestrepo18a}, we estimate a virial factor of $f=1.22\pm0.31$ for H$\beta$. We thus obtain a black hole mass of $\log (M/M_\odot) =7.75\pm0.12$, where the error is only statistical.  Alternatively, the black hole mass can be estimated using the width and luminosity of the broad H$\alpha$ line, to bypass uncertainties in the continuum flux. The luminosity and FWHM of the H$\alpha$ broad component are log~$L_{\rm H\alpha}=41.79\pm0.09$ \ergs\ and $3745\pm205$ \kms, respectively. Thus, using the scaling relation shown in eq.~5 of \citet{reines2013}, we obtained a black hole mass of $\log M/M_\odot=7.67\pm0.06$. These two estimates based on the optical spectral modelling are in good agreement. Combined with the black hole mass estimated above results in a current Eddington ratio of L/L{\rm Edd}=0.4.

\section{Discussion and Conclusions}

We have identified a current type I AGN in a galaxy previously classified as starforming. The candidate was selected by showing optical variations in ZTF light curves, classified as AGN or QSO by the ALeRCE light curve classifier in a parent sample of inactive galaxies with archival optical spectra. The evidence in favour of a newborn AGN is primarily the appearance of prominent, broad Balmer emission lines and a blue continuum on top of the stellar population and narrow emission lines visible in both the old and new spectra. Additional evidence for a transition from inactive to active galactic nucleus are a steady rise in the optical flux for at least 2500 days, since about MJD 55000, as seen in the CRTS light curve, and a change in the MIR color W2-W1, which was consistent with an inactive galaxy around MJD 55500 and later became consistently more AGN-like. 

Additional evidence for current nuclear activity is the detection of a  variable X-ray source, with a 2--10 keV luminosity of about $10^{43}$ erg\,s$^{-1}$. Only one massive star explosion to date, AT2020mrf, has been reported to have reached a similar luminosity, with $L_{0.3-10 \rm{ keV}}\sim 2\times 10^{43}$ erg\,s$^{-1}$\citep{Yao2022}. The X-ray flux of the stellar explosion AT2020mrf, however, declined by an order of magnitude a year later, while ZTF20aaglfpy remained at a similar flux level for the 550 days spanned by the 4 all sky surveys of eROSITA. The optical light curves also differ strongly, with AT2020mrf showing a steep rise and slow decline while ZTF20aaglfpy shows a slow steady rise with recurrent, shorter timescale flaring.

  The narrow emission lines continue to show a ratio indicative of star formation and not AGN. This lack of reaction can be explained if the differential light travel time to the observer via the narrow line region is longer than the time since the ignition of the AGN. We take the difference in time between both spectroscopic observations (i.e. 
  6674 days) as an upper limit for the time between AGN ignition and the observation of the narrow lines in the second spectrum. In this scenario, the differential light travel path should be longer than 18.3 light years or, equivalently, a size larger than 5.6 pc, which is modest compared to typical sizes of NLRs. As a comparison \citet{Bennert2006} determined NLR sizes of 700 pc to over 3 kpc with spatially resolved BPT diagrams of nearby Seyfert galaxies, with transitions to HII-like emission line ratios at similar radii in some objects. 
  
  Broad Balmer lines have been found in a few galaxies with star forming classification based on their narrow line flux ratios, for example by \citet{reines2013}, who focused on dwarf galaxies. Followup spectroscopy of these sources, presented by \citet{Baldassare2016}, showed that all of the broad lines in star forming galaxies either had faded away or were ambiguously detected, pointing to contamination by type II SNe in the spectra with broad lines. Such a stellar explosion scenario is unlikely in the case of ZTF20aaglfpy as discussed above, strengthening the case for a newborn AGN.

{\bf Acknowledgements:}  The authors wish to thank the anonymous referee for their useful suggestions. The authors acknowledge the National Agency for Research and Development (ANID) grants: Millennium Science Initiative Program NCN$19\_058$ and NCN$2023\_002$ (PA, SB, PL, EL, MLM-A) and ICN12\_12009 (PSS,LHG); FONDECYT Regular grants 1201748 (PL) and 1230345 (CR); Programa de Becas/Doctorado Nacional 21212344 (SB), 21200718 (EL); BASAL project FB210003 (CR), the Max-Planck Society through a Partner Group grant (PA, SB), and DLR grant FKZ 50 OR 2307 (MK).

Based on observations obtained with the Samuel Oschin Telescope 48-inch and the 60-inch Telescope at the Palomar Observatory as part of the Zwicky Transient Facility project. ZTF is supported by the National Science Foundation under Grant No. AST-2034437 and a collaboration including Caltech, IPAC, the Weizmann Institute for Science, the Oskar Klein Center at Stockholm University, the University of Maryland, Deutsches Elektronen-Synchrotron and Humboldt University, the TANGO Consortium of Taiwan, the University of Wisconsin at Milwaukee, Trinity College Dublin, Lawrence Livermore National Laboratories, and IN2P3, France. Operations are conducted by COO, IPAC, and UW. The ZTF forced-photometry service was funded under the Heising-Simons Foundation grant 12540303 (PI: Graham). We made use of data products from the Wide-field Infrared Survey Explorer, which is a joint project of the University of California, Los Angeles, and the Jet Propulsion Laboratory/California Institute of Technology, funded by the National Aeronautics and Space Administration; from NEOWISE, which is a project of the Jet Propulsion Laboratory/California Institute of Technology, funded by the Planetary Science Division of the National Aeronautics and Space Administration; from eROSITA aboard SRG, a joint Russian-German science mission supported by the Russian Space Agency (Roskosmos), in the interests of the Russian
Academy of Sciences represented by its Space Research Institute (IKI),
and the Deutsches Zentrum f\"ur Luft- und Raumfahrt (DLR). The SRG
spacecraft was built by Lavochkin Association (NPOL) and its
subcontractors, and is operated by NPOL with support from the Max
Planck Institute for Extraterrestrial Physics (MPE). The development
and construction of the eROSITA X-ray instrument was led by MPE, with
contributions from the Dr. Karl Remeis Observatory Bamberg \& ECAP
(FAU Erlangen-Nuernberg), the University of Hamburg Observatory, the
Leibniz Institute for Astrophysics Potsdam (AIP), and the Institute
for Astronomy and Astrophysics of the University of T\"ubingen, with
the support of DLR and the Max Planck Society. The Argelander
Institute for Astronomy of the University of Bonn and the Ludwig
Maximilians Universit\"at Munich also participated in the science
preparation for eROSITA.

The eROSITA data shown here were processed using the eSASS/NRTA software system developed by the German eROSITA consortium. 
\\

\bibliographystyle{aa} 
\bibliography{bibliography.bib}

\begin{thebibliography}{37}
\expandafter\ifx\csname natexlab\endcsname\relax\def\natexlab#1{#1}\fi

\bibitem[{{Ahumada} {et~al.}(2020){Ahumada}, {Allende Prieto}, {Almeida}, {Anders}, {Anderson}, {Andrews}, {Anguiano}, {Arcodia}, {Armengaud}, {Aubert}, {Avila}, {Avila-Reese}, {Badenes}, {Balland}, {Barger}, {Barrera-Ballesteros}, {Basu}, {Bautista}, {Beaton}, {Beers}, {Benavides}, {Bender}, {Bernardi}, {Bershady}, {Beutler}, {Bidin}, {Bird}, {Bizyaev}, {Blanc}, {Blanton}, {Boquien}, {Borissova}, {Bovy}, {Brandt}, {Brinkmann}, {Brownstein}, {Bundy}, {Bureau}, {Burgasser}, {Burtin}, {Cano-D{\'\i}az}, {Capasso}, {Cappellari}, {Carrera}, {Chabanier}, {Chaplin}, {Chapman}, {Cherinka}, {Chiappini}, {Doohyun Choi}, {Chojnowski}, {Chung}, {Clerc}, {Coffey}, {Comerford}, {Comparat}, {da Costa}, {Cousinou}, {Covey}, {Crane}, {Cunha}, {Ilha}, {Dai}, {Damsted}, {Darling}, {Davidson}, {Davies}, {Dawson}, {De}, {de la Macorra}, {De Lee}, {Queiroz}, {Deconto Machado}, {de la Torre}, {Dell'Agli}, {du Mas des Bourboux}, {Diamond-Stanic}, {Dillon}, {Donor}, {Drory}, {Duckworth}, {Dwelly}, {Ebelke}, {Eftekharzadeh}, {Davis
  Eigenbrot}, {Elsworth}, {Eracleous}, {Erfanianfar}, {Escoffier}, {Fan}, {Farr}, {Fern{\'a}ndez-Trincado}, {Feuillet}, {Finoguenov}, {Fofie}, {Fraser-McKelvie}, {Frinchaboy}, {Fromenteau}, {Fu}, {Galbany}, {Garcia}, {Garc{\'\i}a-Hern{\'a}ndez}, {Garma Oehmichen}, {Ge}, {Geimba Maia}, {Geisler}, {Gelfand}, {Goddy}, {Gonzalez-Perez}, {Grabowski}, {Green}, {Grier}, {Guo}, {Guy}, {Harding}, {Hasselquist}, {Hawken}, {Hayes}, {Hearty}, {Hekker}, {Hogg}, {Holtzman}, {Horta}, {Hou}, {Hsieh}, {Huber}, {Hunt}, {Ider Chitham}, {Imig}, {Jaber}, {Jimenez Angel}, {Johnson}, {Jones}, {J{\"o}nsson}, {Jullo}, {Kim}, {Kinemuchi}, {Kirkpatrick}, {Kite}, {Klaene}, {Kneib}, {Kollmeier}, {Kong}, {Kounkel}, {Krishnarao}, {Lacerna}, {Lan}, {Lane}, {Law}, {Le Goff}, {Leung}, {Lewis}, {Li}, {Lian}, {Lin}, {Long}, {Longa-Pe{\~n}a}, {Lundgren}, {Lyke}, {Mackereth}, {MacLeod}, {Majewski}, {Manchado}, {Maraston}, {Martini}, {Masseron}, {Masters}, {Mathur}, {McDermid}, {Merloni}, {Merrifield}, {M{\'e}sz{\'a}ros}, {Miglio}, {Minniti},
  {Minsley}, {Miyaji}, {Mohammad}, {Mosser}, {Mueller}, {Muna}, {Mu{\~n}oz-Guti{\'e}rrez}, {Myers}, {Nadathur}, {Nair}, {Nandra}, {Correa do Nascimento}, {Nevin}, {Newman}, {Nidever}, {Nitschelm}, {Noterdaeme}, {O'Connell}, {Olmstead}, {Oravetz}, {Oravetz}, {Osorio}, {Pace}, {Padilla}, {Palanque-Delabrouille}, {Palicio}, {Pan}, {Pan}, {Parker}, {Paviot}, {Peirani}, {Ram{\'r}ez}, {Penny}, {Percival}, {Perez-Fournon}, {P{\'e}rez-R{\`a}fols}, {Petitjean}, {Pieri}, {Pinsonneault}, {Poovelil}, {Povick}, {Prakash}, {Price-Whelan}, {Raddick}, {Raichoor}, {Ray}, {Rembold}, {Rezaie}, {Riffel}, {Riffel}, {Rix}, {Robin}, {Roman-Lopes}, {Rom{\'a}n-Z{\'u}{\~n}iga}, {Rose}, {Ross}, {Rossi}, {Rowlands}, {Rubin}, {Salvato}, {S{\'a}nchez}, {S{\'a}nchez-Menguiano}, {S{\'a}nchez-Gallego}, {Sayres}, {Schaefer}, {Schiavon}, {Schimoia}, {Schlafly}, {Schlegel}, {Schneider}, {Schultheis}, {Schwope}, {Seo}, {Serenelli}, {Shafieloo}, {Shamsi}, {Shao}, {Shen}, {Shetrone}, {Shirley}, {Silva Aguirre}, {Simon}, {Skrutskie}, {Slosar},
  {Smethurst}, {Sobeck}, {Sodi}, {Souto}, {Stark}, {Stassun}, {Steinmetz}, {Stello}, {Stermer}, {Storchi-Bergmann}, {Streblyanska}, {Stringfellow}, {Stutz}, {Su{\'a}rez}, {Sun}, {Taghizadeh-Popp}, {Talbot}, {Tayar}, {Thakar}, {Theriault}, {Thomas}, {Thomas}, {Tinker}, {Tojeiro}, {Toledo}, {Tremonti}, {Troup}, {Tuttle}, {Unda-Sanzana}, {Valentini}, {Vargas-Gonz{\'a}lez}, {Vargas-Maga{\~n}a}, {V{\'a}zquez-Mata}, {Vivek}, {Wake}, {Wang}, {Weaver}, {Weijmans}, {Wild}, {Wilson}, {Wilson}, {Wolthuis}, {Wood-Vasey}, {Yan}, {Yang}, {Y{\`e}che}, {Zamora}, {Zarrouk}, {Zasowski}, {Zhang}, {Zhao}, {Zhao}, {Zheng}, {Zheng}, {Zhu}, \& {Zou}}]{Ahumada2020}
{Ahumada}, R., {Allende Prieto}, C., {Almeida}, A., {et~al.} 2020, \apjs, 249, 3

\bibitem[{{Baldassare} {et~al.}(2016){Baldassare}, {Reines}, {Gallo}, {Greene}, {Graur}, {Geha}, {Hainline}, {Carroll}, \& {Hickox}}]{Baldassare2016}
{Baldassare}, V.~F., {Reines}, A.~E., {Gallo}, E., {et~al.} 2016, \apj, 829, 57

\bibitem[{{Baldwin} \& {Netzer}(1978)}]{Baldwin1978}
{Baldwin}, J.~A. \& {Netzer}, H. 1978, \apj, 226, 1

\bibitem[{Baldwin {et~al.}(1981)Baldwin, Phillips, \& Terlevich}]{baldwin1981classification}
Baldwin, J.~A., Phillips, M.~M., \& Terlevich, R. 1981, Publications of the Astronomical Society of the Pacific, 93, 5

\bibitem[{{Bennert} {et~al.}(2006){Bennert}, {Jungwiert}, {Komossa}, {Haas}, \& {Chini}}]{Bennert2006}
{Bennert}, N., {Jungwiert}, B., {Komossa}, S., {Haas}, M., \& {Chini}, R. 2006, \aap, 459, 55

\bibitem[{{Bentz} {et~al.}(2013){Bentz}, {Denney}, {Grier}, {Barth}, {Peterson}, {Vestergaard}, {Bennert}, {Canalizo}, {De Rosa}, {Filippenko}, {Gates}, {Greene}, {Li}, {Malkan}, {Pogge}, {Stern}, {Treu}, \& {Woo}}]{bentz13}
{Bentz}, M.~C., {Denney}, K.~D., {Grier}, C.~J., {et~al.} 2013, \apj, 767, 149

\bibitem[{{Brogan} {et~al.}(2023){Brogan}, {Krumpe}, {Homan}, {Urrutia}, {Granzer}, {Husemann}, {Neumann}, {Gaspari}, {Vaughan}, {Croom}, {Combes}, {P{\'e}rez Torres}, {Coil}, {McElroy}, {Winkel}, \& {Singha}}]{Brogan2023}
{Brogan}, R., {Krumpe}, M., {Homan}, D., {et~al.} 2023, \aap, 677, A116

\bibitem[{{Brunner} {et~al.}(2022){Brunner}, {Liu}, {Lamer}, {Georgakakis}, {Merloni}, {Brusa}, {Bulbul}, {Dennerl}, {Friedrich}, {Liu}, {Maitra}, {Nandra}, {Ramos-Ceja}, {Sanders}, {Stewart}, {Boller}, {Buchner}, {Clerc}, {Comparat}, {Dwelly}, {Eckert}, {Finoguenov}, {Freyberg}, {Ghirardini}, {Gueguen}, {Haberl}, {Kreykenbohm}, {Krumpe}, {Osterhage}, {Pacaud}, {Predehl}, {Reiprich}, {Robrade}, {Salvato}, {Santangelo}, {Schrabback}, {Schwope}, \& {Wilms}}]{2022brunner}
{Brunner}, H., {Liu}, T., {Lamer}, G., {et~al.} 2022, \aap, 661, A1

\bibitem[{{Cappellari}(2017)}]{Cappellari2017}
{Cappellari}, M. 2017, MNRAS, 466, 798

\bibitem[{{Cappellari} \& {Emsellem}(2004)}]{2004PASP..116..138C}
{Cappellari}, M. \& {Emsellem}, E. 2004, \pasp, 116, 138

\bibitem[{{Chilingarian} \& {Zolotukhin}(2012)}]{Chilingarian2012}
{Chilingarian}, I.~V. \& {Zolotukhin}, I.~Y. 2012, \mnras, 419, 1727

\bibitem[{{Chilingarian} {et~al.}(2017){Chilingarian}, {Zolotukhin}, {Katkov}, {Melchior}, {Rubtsov}, \& {Grishin}}]{Chilingarian2017}
{Chilingarian}, I.~V., {Zolotukhin}, I.~Y., {Katkov}, I.~Y., {et~al.} 2017, \apjs, 228, 14

\bibitem[{{Comparat} {et~al.}(2020){Comparat}, {Merloni}, {Dwelly}, {Salvato}, {Schwope}, {Coffey}, {Wolf}, {Arcodia}, {Liu}, {Buchner}, {Nandra}, {Georgakakis}, {Clerc}, {Brusa}, {Brownstein}, {Schneider}, {Pan}, \& {Bizyaev}}]{Comparat2020}
{Comparat}, J., {Merloni}, A., {Dwelly}, T., {et~al.} 2020, \aap, 636, A97

\bibitem[{{Drake} {et~al.}(2009){Drake}, {Djorgovski}, {Mahabal}, {Beshore}, {Larson}, {Graham}, {Williams}, {Christensen}, {Catelan}, {Boattini}, {Gibbs}, {Hill}, \& {Kowalski}}]{Drake2009}
{Drake}, A.~J., {Djorgovski}, S.~G., {Mahabal}, A., {et~al.} 2009, \apj, 696, 870

\bibitem[{{Duras} {et~al.}(2020){Duras}, {Bongiorno}, {Ricci}, {Piconcelli}, {Shankar}, {Lusso}, {Bianchi}, {Fiore}, {Maiolino}, {Marconi}, {Onori}, {Sani}, {Schneider}, {Vignali}, \& {La Franca}}]{Duras2020}
{Duras}, F., {Bongiorno}, A., {Ricci}, F., {et~al.} 2020, \aap, 636, A73

\bibitem[{Falc{\'o}n-Barroso {et~al.}(2011)Falc{\'o}n-Barroso, S{\'a}nchez-Bl{\'a}zquez, Vazdekis, Ricciardelli, Cardiel, Cenarro, Gorgas, \& Peletier}]{falcon2011updated}
Falc{\'o}n-Barroso, J., S{\'a}nchez-Bl{\'a}zquez, P., Vazdekis, A., {et~al.} 2011, Astronomy \& Astrophysics, 532, A95

\bibitem[{{F{\"o}rster} {et~al.}(2021){F{\"o}rster}, {Cabrera-Vives}, {Castillo-Navarrete}, {Est{\'e}vez}, {S{\'a}nchez-S{\'a}ez}, {Arredondo}, {Bauer}, {Carrasco-Davis}, {Catelan}, {Elorrieta}, {Eyheramendy}, {Huijse}, {Pignata}, {Reyes}, {Reyes}, {Rodr{\'\i}guez-Mancini}, {Ruz-Mieres}, {Valenzuela}, {{\'A}lvarez-Maldonado}, {Astorga}, {Borissova}, {Clocchiatti}, {De Cicco}, {Donoso-Oliva}, {Hern{\'a}ndez-Garc{\'\i}a}, {Graham}, {Jord{\'a}n}, {Kurtev}, {Mahabal}, {Maureira}, {Mu{\~n}oz-Arancibia}, {Molina-Ferreiro}, {Moya}, {Palma}, {P{\'e}rez-Carrasco}, {Protopapas}, {Romero}, {Sabatini-Gacitua}, {S{\'a}nchez}, {San Mart{\'\i}n}, {Sep{\'u}lveda-Cobo}, {Vera}, \& {Vergara}}]{Forster21}
{F{\"o}rster}, F., {Cabrera-Vives}, G., {Castillo-Navarrete}, E., {et~al.} 2021, \aj, 161, 242

\bibitem[{{Hickox} {et~al.}(2014){Hickox}, {Mullaney}, {Alexander}, {Chen}, {Civano}, {Goulding}, \& {Hainline}}]{Hickox2014}
{Hickox}, R.~C., {Mullaney}, J.~R., {Alexander}, D.~M., {et~al.} 2014, \apj, 782, 9

\bibitem[{{Ichikawa} {et~al.}(2019){Ichikawa}, {Ueda}, {Bae}, {Kawamuro}, {Matsuoka}, {Toba}, \& {Shidatsu}}]{Ichikawa2019}
{Ichikawa}, K., {Ueda}, J., {Bae}, H.-J., {et~al.} 2019, \apj, 870, 65

\bibitem[{Kewley {et~al.}(2001)Kewley, Dopita, Sutherland, Heisler, \& Trevena}]{kewley2001theoretical}
Kewley, L.~J., Dopita, M., Sutherland, R., Heisler, C., \& Trevena, J. 2001, The Astrophysical Journal, 556, 121

\bibitem[{Kewley {et~al.}(2006)Kewley, Groves, Kauffmann, \& Heckman}]{kewley2006host}
Kewley, L.~J., Groves, B., Kauffmann, G., \& Heckman, T. 2006, Monthly Notices of the Royal Astronomical Society, 372, 961

\bibitem[{{L{\'o}pez-Navas} {et~al.}(2023){L{\'o}pez-Navas}, {Ar{\'e}valo}, {Bernal}, {Graham}, {Hern{\'a}ndez-Garc{\'\i}a}, {Lira}, \& {S{\'a}nchez-S{\'a}ez}}]{Lopez23}
{L{\'o}pez-Navas}, E., {Ar{\'e}valo}, P., {Bernal}, S., {et~al.} 2023, \mnras, 518, 1531

\bibitem[{{MacLeod} {et~al.}(2010){MacLeod}, {Ivezi{\'c}}, {Kochanek}, {Koz{\l}owski}, {Kelly}, {Bullock}, {Kimball}, {Sesar}, {Westman}, {Brooks}, {Gibson}, {Becker}, \& {de Vries}}]{MacLeod10}
{MacLeod}, C.~L., {Ivezi{\'c}}, {\v Z}., {Kochanek}, C.~S., {et~al.} 2010, \apj, 721, 1014

\bibitem[{{Masci} {et~al.}(2019){Masci}, {Laher}, {Rusholme}, {Shupe}, {Groom}, {Surace}, {Jackson}, {Monkewitz}, {Beck}, {Flynn}, {Terek}, {Landry}, {Hacopians}, {Desai}, {Howell}, {Brooke}, {Imel}, {Wachter}, {Ye}, {Lin}, {Cenko}, {Cunningham}, {Rebbapragada}, {Bue}, {Miller}, {Mahabal}, {Bellm}, {Patterson}, {Juri{\'c}}, {Golkhou}, {Ofek}, {Walters}, {Graham}, {Kasliwal}, {Dekany}, {Kupfer}, {Burdge}, {Cannella}, {Barlow}, {Van Sistine}, {Giomi}, {Fremling}, {Blagorodnova}, {Levitan}, {Riddle}, {Smith}, {Helou}, {Prince}, \& {Kulkarni}}]{Masci19}
{Masci}, F.~J., {Laher}, R.~R., {Rusholme}, B., {et~al.} 2019, \pasp, 131, 018003

\bibitem[{Meisner {et~al.}(2023)Meisner, Caselden, Schlafly, \& Kiwy}]{Meisner2023}
Meisner, A.~M., Caselden, D., Schlafly, E.~F., \& Kiwy, F. 2023, The Astronomical Journal, 165, 36

\bibitem[{{Mej{\'{\i}}a-Restrepo} {et~al.}(2018){Mej{\'{\i}}a-Restrepo}, {Lira}, {Netzer}, {Trakhtenbrot}, \& {Capellupo}}]{MejiaRestrepo18a}
{Mej{\'{\i}}a-Restrepo}, J.~E., {Lira}, P., {Netzer}, H., {Trakhtenbrot}, B., \& {Capellupo}, D.~M. 2018, Nature Astronomy, 2, 63

\bibitem[{{Predehl} {et~al.}(2021){Predehl}, {Andritschke}, {Arefiev}, {Babyshkin}, {Batanov}, {Becker}, {B{\"o}hringer}, {Bogomolov}, {Boller}, {Borm}, {Bornemann}, {Br{\"a}uninger}, {Br{\"u}ggen}, {Brunner}, {Brusa}, {Bulbul}, {Buntov}, {Burwitz}, {Burkert}, {Clerc}, {Churazov}, {Coutinho}, {Dauser}, {Dennerl}, {Doroshenko}, {Eder}, {Emberger}, {Eraerds}, {Finoguenov}, {Freyberg}, {Friedrich}, {Friedrich}, {F{\"u}rmetz}, {Georgakakis}, {Gilfanov}, {Granato}, {Grossberger}, {Gueguen}, {Gureev}, {Haberl}, {H{\"a}lker}, {Hartner}, {Hasinger}, {Huber}, {Ji}, {Kienlin}, {Kink}, {Korotkov}, {Kreykenbohm}, {Lamer}, {Lomakin}, {Lapshov}, {Liu}, {Maitra}, {Meidinger}, {Menz}, {Merloni}, {Mernik}, {Mican}, {Mohr}, {M{\"u}ller}, {Nandra}, {Nazarov}, {Pacaud}, {Pavlinsky}, {Perinati}, {Pfeffermann}, {Pietschner}, {Ramos-Ceja}, {Rau}, {Reiffers}, {Reiprich}, {Robrade}, {Salvato}, {Sanders}, {Santangelo}, {Sasaki}, {Scheuerle}, {Schmid}, {Schmitt}, {Schwope}, {Shirshakov}, {Steinmetz}, {Stewart}, {Str{\"u}der},
  {Sunyaev}, {Tenzer}, {Tiedemann}, {Tr{\"u}mper}, {Voron}, {Weber}, {Wilms}, \& {Yaroshenko}}]{2021erosita}
{Predehl}, P., {Andritschke}, R., {Arefiev}, V., {et~al.} 2021, \aap, 647, A1

\bibitem[{{Reines} {et~al.}(2013){Reines}, {Greene}, \& {Geha}}]{reines2013}
{Reines}, A.~E., {Greene}, J.~E., \& {Geha}, M. 2013, \apj, 775, 116

\bibitem[{{S{\'a}nchez-S{\'a}ez} {et~al.}(2018){S{\'a}nchez-S{\'a}ez}, {Lira}, {Mej{\'{\i}}a-Restrepo}, {Ho}, {Ar{\'e}valo}, {Kim}, {Cartier}, \& {Coppi}}]{Sanchez-Saez18}
{S{\'a}nchez-S{\'a}ez}, P., {Lira}, P., {Mej{\'{\i}}a-Restrepo}, J., {et~al.} 2018, \apj, 864, 87

\bibitem[{{S{\'a}nchez-S{\'a}ez} {et~al.}(2021){S{\'a}nchez-S{\'a}ez}, {Reyes}, {Valenzuela}, {F{\"o}rster}, {Eyheramendy}, {Elorrieta}, {Bauer}, {Cabrera-Vives}, {Est{\'e}vez}, {Catelan}, {Pignata}, {Huijse}, {De Cicco}, {Ar{\'e}valo}, {Carrasco-Davis}, {Abril}, {Kurtev}, {Borissova}, {Arredondo}, {Castillo-Navarrete}, {Rodriguez}, {Ruz-Mieres}, {Moya}, {Sabatini-Gacit{\'u}a}, {Sep{\'u}lveda-Cobo}, \& {Camacho-I{\~n}iguez}}]{Sanchez-Saez21}
{S{\'a}nchez-S{\'a}ez}, P., {Reyes}, I., {Valenzuela}, C., {et~al.} 2021, \aj, 161, 141

\bibitem[{{Schawinski} {et~al.}(2007){Schawinski}, {Thomas}, {Sarzi}, {Maraston}, {Kaviraj}, {Joo}, {Yi}, \& {Silk}}]{schawinsky2007}
{Schawinski}, K., {Thomas}, D., {Sarzi}, M., {et~al.} 2007, \mnras, 382, 1415

\bibitem[{{Schulze} \& {Wisotzki}(2010)}]{Schulze2010}
{Schulze}, A. \& {Wisotzki}, L. 2010, \aap, 516, A87

\bibitem[{{Shankar} {et~al.}(2013){Shankar}, {Weinberg}, \& {Miralda-Escud{\'e}}}]{Shankar2013}
{Shankar}, F., {Weinberg}, D.~H., \& {Miralda-Escud{\'e}}, J. 2013, \mnras, 428, 421

\bibitem[{{Shen} {et~al.}(2008){Shen}, {Greene}, {Strauss}, {Richards}, \& {Schneider}}]{Shen08}
{Shen}, Y., {Greene}, J.~E., {Strauss}, M.~A., {Richards}, G.~T., \& {Schneider}, D.~P. 2008, \apj, 680, 169

\bibitem[{{Sun} {et~al.}(2015){Sun}, {Trump}, {Brandt}, {Luo}, {Alexander}, {Jahnke}, {Rosario}, {Wang}, \& {Xue}}]{Sun2015}
{Sun}, M., {Trump}, J.~R., {Brandt}, W.~N., {et~al.} 2015, \apj, 802, 14

\bibitem[{{Wright} {et~al.}(2010){Wright}, {Eisenhardt}, {Mainzer}, {Ressler}, {Cutri}, {Jarrett}, {Kirkpatrick}, {Padgett}, {McMillan}, {Skrutskie}, {Stanford}, {Cohen}, {Walker}, {Mather}, {Leisawitz}, {Gautier}, {McLean}, {Benford}, {Lonsdale}, {Blain}, {Mendez}, {Irace}, {Duval}, {Liu}, {Royer}, {Heinrichsen}, {Howard}, {Shannon}, {Kendall}, {Walsh}, {Larsen}, {Cardon}, {Schick}, {Schwalm}, {Abid}, {Fabinsky}, {Naes}, \& {Tsai}}]{Wright2010}
{Wright}, E.~L., {Eisenhardt}, P. R.~M., {Mainzer}, A.~K., {et~al.} 2010, \aj, 140, 1868

\bibitem[{Yao {et~al.}(2022)Yao, Ho, Medvedev, J., Perley, Kulkarni, Chandra, Sazonov, Gilfanov, Khorunzhev, Khatami, \& Sunyaev}]{Yao2022}
Yao, Y., Ho, A. Y.~Q., Medvedev, P., {et~al.} 2022, The Astrophysical Journal, 934, 104

\end{thebibliography}

\appendix
\section{Considerations on the different apertures for old and new spectra}
We note that the SDSS spectrum was taken with a 3$\arcsec$ diameter circular fiber so it included more starlight than the 1$\arcsec$-wide slit spectrum taken with SOAR. To check whether this difference could have hidden the broad emission lines we re-extracted the SOAR spectrum from a region of 3$\arcsec$ along the spatial axis, removing the flux from the central 1$\arcsec$, to sample the spectra in the "missing" outer regions. Adding three times this external spectrum to the central 1$\arcsec$ spectrum simulates what would be obtained from the SDSS 3$\arcsec$ fiber, even overcorrecting for the missing starlight. This experiment still resulted in very significant broad H$\alpha$ line, proving that additional starlight in the SDSS spectrum was not hiding broad lines emission. Fig.\ref{fig:comparison} shows the result of this experiment. 
\begin{figure}
    \centering
    \includegraphics[width=0.49\textwidth]{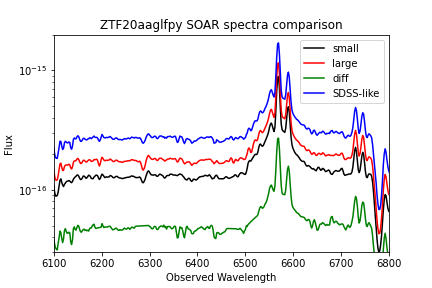}
    \caption{Different extractions of the new, SOAR spectrum of ZF20aaglfpy. The spectrum extracted along the slit with a 1$\arcsec$ length along the spatial axis, labeled "small" in the figure, has only a slightly stronger broad H$_\alpha$ line than the spectrum extracted from a longer, 3$\arcsec$  region (labeled "large"), and is very similar to the spectrum of the difference between both (labeled "diff"). The "SDSS-like" spectrum is the sum of "large" plus two times "diff" and should account for all the light in the 3$\arcsec$ aperture of the SDSS spectrum, even including additional starlight. The broad component in H$_\alpha$ is still very significant in this compound spectrum.  }
    \label{fig:comparison}
\end{figure}

\end{document}